\documentclass[aps,prb,twocolumn,prabib,amsmath,amssymb,superscriptaddress,notitlepage]{revtex4-1}
\usepackage{graphics}
\usepackage{bm}
\usepackage{dcolumn}
\usepackage{natbib}
\usepackage{epstopdf}
\usepackage{float}
\usepackage{graphicx}
\usepackage{epsfig}
\usepackage[pdfstartview=FitH]{hyperref}
\usepackage{color}
\usepackage{appendix}
\usepackage{ulem}

\newcommand{\rr}{{\bf r}}

\begin{document}
\title{SU$(N)$ fractional quantum Hall effects in topological flat bands}
\author{Tian-Sheng Zeng}
\affiliation{Department of Physics, University of Texas at Dallas, Richardson, Texas 75080, USA}
\affiliation{Department of Physics and Astronomy, California State University, Northridge, California 91330, USA}
\author{D. N. Sheng}
\affiliation{Department of Physics and Astronomy, California State University, Northridge, California 91330, USA}
\date{\today}
\begin{abstract}
We study $N$-component interacting particles (hardcore bosons and fermions) loaded in topological lattice models with SU$(N)$-invariant interactions based on exact diagonalization and density matrix renormalization group method. By tuning the interplay of interspecies and intraspecies interactions, we demonstrate that a class of SU$(N)$ fractional quantum Hall states can emerge at fractional filling factors $\nu=N/(N+1)$ for bosons ($\nu=N/(2N+1)$ for fermions) in the lowest Chern band, characterized by the
nontrivial fractional Hall responses and the fractional charge pumping. Moreover, we establish a topological characterization based on the $\mathbf{K}$ matrix, and discuss the close relationship to the fractional quantum Hall physics in topological flat bands with Chern number $N$.
\end{abstract}
\maketitle
\section{introduction}

Since the theoretical discovery of fractionalized topological ordered phases in topological flat bands with Chern number one~\cite{Sun2011,Neupert2011,Sheng2011,Tang2011,Wang2011,Regnault2011}, the study of fractional Chern insulator (FCI) has became an exciting  subfield of condensed matter physics.
In analogy to the Laughlin fractional quantum Hall (FQH) states in two-dimensional Landau levels~\cite{BL2013}, recent numerical studies suggest that a rich series of Abelian FCI emerges when single-component particles partially occupy topological flat bands with higher Chern number $C>1$ at fillings $\nu=1/(MC+1)$ ($M=1$ for hardcore bosons and for $M=2$ spinless fermions)~\cite{LBFL2012,Wang2012r,Yang2012,Sterdyniak2013,Wang2013,Moller2015,Sterdyniak2015b}. For $C=2$, these FCIs are believed to be color-entangled lattice versions of two-component Halperin $(mmn)$ FQH states~\cite{Halperin1983}, and the corresponding Haldane pseudopotential Hamiltonians for these FCIs can be constructed~\cite{Barkeshli2012,Wu2013,Wu2014}. For two-component systems, by tuning interspecies interaction, one would expect a rich class of spin-singlet FQH states~\cite{Balatsky1991,Lopez1995,Read1997,Ardonne1999,Lopez2001,Ardonne2002,Furukawa2012,Grass2012,YHWu2013,Grass2014}, and the tunneling-induced transition from the Halperin (332) state to a non-Abelian phase~\cite{Papic2010}. Here we refer the different components of the system as the (pseudo) spin degrees of freedom.

Experimentally, the zeroth Landau level of monolayer graphene which hosts an approximate SU$(4)$ pseudospin symmetry in the presence of spin and valley degrees of freedom, provides a door to exploring the SU$(4)$ symmetric FQH states or other competing broken symmetry states at partial filling under the interplay between electronic correlations and the inherent symmetries of graphene~\cite{Bolotin2009,Du2009,Dean2011,Young2012,Feldman2012,Feldman2013}. The SU$(4)$ generalizations of Halperin's wavefunctions characterized by the integer valued symmetric $\mathbf{K}$ matrix, to these SU$(4)$ FQH states in graphene sheets, are proposed in Refs.~\cite{Toke2007,Goerbig2007,Gail2008,Modak2011}, while the SU$(3)$ generalizations to bosonic FQH states are also studied in Refs.~\cite{Sterdyniak2015b,Reijnders2002,Reijnders2004}, and a comprehensive phase diagram of FQH states in multicomponent systems is constructed using composite fermion theory~\cite{Balram2015}. Under the high Zeeman effect terms, the four-fold degenerate levels are broken, and SU$(2)$ symmetry can be restored. Thus one would recover the picture of fully spin polarized FQH states in the SU$(2)$ valley space~\cite{Toke2012,Abanin2013,Sodemann2014,Jolicoeur2014}. In cold atom physics the SU$(2)$ symmetric Haldane-honeycomb insulator has also been achieved with two-component fermionic $^{40}$K atoms~\cite{Jotzu2014}, which is also promising for hosting $(N>2)$-component systems. More excitingly, the fractional Chern insulators are experimentally observed in a bilayer graphene heterostructure~\cite{Spanton2017}. These related experimental advances would enable the study of the $N$-component quantum Hall effect in topological flat band models.

To understand the topological nature of the multicomponent systems, the integer valued symmetric $\mathbf{K}$ matrix, which classifies the topological order at different fillings for Abelian multicomponent systems according to the Chern-Simons theory~\cite{Wen1992a,Wen1992b,Blok1990,Blok1991}, has been numerically demonstrated recently from the inverse of the Chern number matrix in Ref.~\cite{Zeng2017}, where the topological information from $\mathbf{K}$ matrix of two-component Halperin (221) and (331) states are discussed. A systematic characterization of the topological nature of FCI for $C>2$ or multicomponent ($N>2$) systems remains absent, especially so far there are no studies of the $\mathbf{K}$ matrix for such systems, which is the focus of the present work.

In this paper, we systematically study the FQH physics for $N$-component particles with SU$(N)$-invariant interactions in several microscopic topological lattice models with $N$ up to four. Through exact diagonalization (ED) and density matrix renormalization group (DMRG) methods, we show that for a given fractional filling factor, a class of incompressible FQH states emerge under the interplay of interaction and band topology. Topological properties of these states are  characterized by the $\mathbf{K}$ matrix~\cite{Blok1990}, including (\textrm{i}) fractional quantized topological invariants related to Hall conductances, and (\textrm{ii}) degenerate ground states manifold and fractional charge pumping
under the adiabatic insertion of flux quantum.

This paper is organized as follows. In Sec.~\ref{model}, we introduce the SU$(N)$ symmetric Hamiltonian of $N$-component particles loaded on two types of topological lattice models as $\pi$-flux checkerboard and Haldane-honeycomb lattices, and give a description of our numerical methods. In Sec.~\ref{kmatrix}, we study the many-body ground states of these $N$-component particles in the strong interaction regime, present numerical results of the $\mathbf{K}$ matrix by exact diagonalization at fillings $\nu=N/(N+1)$ for hardcore bosons and $\nu=N/(2N+1)$ for fermions. In Sec.~\ref{pumping}, we calculate the fractional charge pumping of the FQH states from adiabatic DMRG, and demonstrate the quantized drag Hall conductance. In Sec.~\ref{fci}, we discuss the close relationship between these $N$-component FQH states and the physics in topological flat bands with Chern number $N$. Finally, in Sec.~\ref{summary}, we summarize our results and discuss the prospect of investigating nontrivial topological states in multicomponent quantum gases.

\section{Models and Methods}\label{model}

\begin{figure}[t]
  \includegraphics[height=1.6in,width=3.4in]{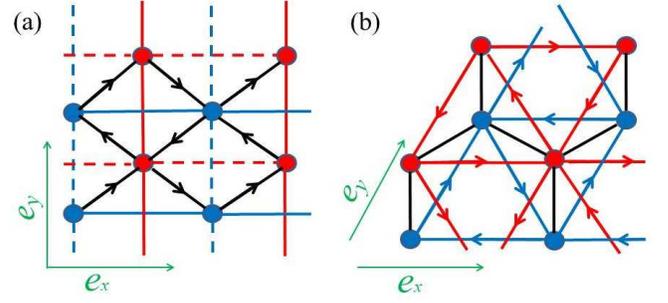}
  \caption{\label{lattice}(Color online) (a) The $\pi$-flux checkerboard lattice model in Eq.~\ref{cb} and (b) the Haldane-honeycomb lattice model in Eq.~\ref{hc}. The arrow directions present the signs of the phases $\phi$ in the hopping terms. The hopping between sites in the sublattice is labeled by the blue (or red) color respectively, while the hopping between different sublattice sites is labeled by the black color. For the checkerboard lattice, the next-nearest-neighbor hopping amplitudes are $t_{\rr,\rr'}'=\pm t'$ along the solid (dotted) lines. $e_{x,y}$ indicate the real-space lattice translational vectors.}
\end{figure}

We consider the following SU$(N)$ symmetric Hamiltonian of interacting $N$-component particles (hardcore bosons or fermions) with SU$(N)$-invariant interactions in two typical topological lattice models,
\begin{align}
  \quad\quad\quad&H=\sum_{\sigma}H_{CB}^{\sigma}+V_{int},\label{cbl}\\
  \quad\quad\quad&H=\sum_{\sigma}H_{HC}^{\sigma}+V_{int},\\
  V_{int}=U&\sum_{\sigma\neq\sigma'}\sum_{\rr}n_{\rr,\sigma}n_{\rr,\sigma'}+V\sum_{\sigma,\sigma'}\sum_{\langle\rr,\rr'\rangle}n_{\rr',\sigma}n_{\rr,\sigma'},\label{interact}
\end{align}
where $H_{CB}^{\sigma}$ denotes the particle hopping terms of the $\sigma$-th component $\sigma=1,2,\cdots,N$ in the $\pi$-flux checkerboard (CB) lattice plotted in Fig.~\ref{lattice}(a),
\begin{align}
  &H_{CB}^{\sigma}=-t\!\sum_{\langle\rr,\rr'\rangle}\!\big[c_{\rr',\sigma}^{\dag}c_{\rr,\sigma}\exp(i\phi_{\rr'\rr})+H.c.\big]\nonumber\\
  &-\!\sum_{\langle\langle\rr,\rr'\rangle\rangle}\!\!\! t_{\rr,\rr'}'c_{\rr',\sigma}^{\dag}c_{\rr,\sigma}
  -t''\!\sum_{\langle\langle\rr,\rr'\rangle\rangle}\!\!\!\! c_{\rr',\sigma}^{\dag}c_{\rr,\sigma}+H.c.,\label{cb}
\end{align}
and $H_{HC}^{\sigma}$ the particle hopping terms of the $\sigma$-th component $\sigma=1,2,\cdots,N$ in Haldane-honeycomb (HC) lattice plotted in Fig.~\ref{lattice}(b),
\begin{align}
  &H_{HC}^{\sigma}=-t'\!\sum_{\langle\langle\rr,\rr'\rangle\rangle}[c_{\rr',\sigma}^{\dag}c_{\rr,\sigma}\exp(i\phi_{\rr'\rr})+H.c.]\nonumber\\
  &-t\!\sum_{\langle\rr,\rr'\rangle}\!\! c_{\rr',\sigma}^{\dag}c_{\rr,\sigma}
  -t''\!\sum_{\langle\langle\rr,\rr'\rangle\rangle}\!\!\!\! c_{\rr',\sigma}^{\dag}c_{\rr,\sigma}+H.c..\label{hc}
\end{align}
Here $c_{\rr,\sigma}^{\dag}$ is the particle creation operator of the $\sigma$-th component at site $\rr$, $n_{\rr,\sigma}=c_{\rr,\sigma}^{\dag}c_{\rr,\sigma}$ is the particle number operator of the $\sigma$-th component at site $\rr$, $\langle\ldots\rangle$,$\langle\langle\ldots\rangle\rangle$ and $\langle\langle\langle\ldots\rangle\rangle\rangle$ denote the nearest-neighbor, the next-nearest-neighbor, and the next-next-nearest-neighbor pairs of sites, respectively. In this work, we consider the on-site interspecies and nearest neighboring intraspecies interactions in Eq.~\ref{interact}. In the flat band limit, we take the parameters $t'=0.3t,t''=-0.2t,\phi=\pi/4$ for checkerboard lattice, as in Ref.~\cite{Zeng2015}, while $t'=0.6t,t''=-0.58t,\phi=2\pi/5$ for honeycomb lattice, as in Refs.~\cite{Wang2011,Wang2012}. $U$ is the strength of the onsite interspecies interaction while $V$ is the strength of nearest-neighbor particle correlations in topological flat bands, playing the analogous role of Haldane pseudopotentials for multicomponent FQHE system in Landau levels~\cite{Peterson2010,Davenport2012,Lee2015}. Here we would numerically address the emergence of a series of $N$-component FQH states in topological flat band models with SU$(N)$-invariant interactions, where ``$N$-component'' serves as a generic label for spin or pseudospin (layer, subband, or valley) degrees of freedom. In cold atomic systems, such high SU$(N=2F+1)$ symmetry can be explored using alkali or alkaline-earth atoms~\cite{Wu2003,Wu2012b} in the lowest hyperfine or nuclear spin multiplets $|F,M\rangle$ with a tunable Hubbard repulsion, such as fermion SU(6) $^{173}$Yb with a nuclear spin $I=5/2$ and bosonic SU(3) $^{87}$Rb with a hyperfine spin $F=1$. The interactions among the different internal atomic states are SU$(N)$-symmetric.

In the ED study, we explore the many-body ground state of $H$ in a finite system of $N_x\times N_y$ unit cells (the total number of sites is $N_s=q\times N_x\times N_y$, with $q$ the number of inequivalent sites within a unit cell.) The total filling of the lowest Chern band is $\nu=qN_e/N_s$, where $N_e=\sum_{\sigma}N_{\sigma}$ is the total particle number with global $U(1)$-symmetry. With the translational symmetry, the energy states are labeled by the total momentum $K=(K_x,K_y)$ in units of $(2\pi/N_x,2\pi/N_y)$ in the Brillouin zone. While the ED calculations on the periodic lattice are limited to a small system, we exploit DMRG for larger systems on cylinder geometry with open boundary conditions in the $x$-direction and periodic boundary conditions in the $y$-direction. We keep the number of states 1200--2400 to obtain accurate results for different system sizes, and the maximal discarded truncation error is less than $10^{-5}$. To avoid the local minimum state, we choose different random initial states with the sweep number more than 20 to get the most convergent ground state.

\section{Multicomponent FQH states}\label{kmatrix}

In this section, we begin with discussing the emergence of multicomponent FQH states at a given filling $\nu=N/(N+1)$ for hardcore bosons and $\nu=N/(2N+1)$ for fermions, respectively. For generic Abelian $N$-component FQH systems at filling $\nu$, they can be classified by a class of the integer valued symmetric $\mathbf{K}$ matrix of the rank $N$~\cite{Wen1992a,Wen1992b,Blok1990,Blok1991}. The ground state degeneracy of FQH systems is given by the determinant $\det\mathbf{K}$, and the number of chiral edge modes is determined by the sign of the eigenvalues of $\mathbf{K}$ matrix. The $\mathbf{K}$ matrix is related to their Hall conductance (Chern number matrix $\mathbf{C}$ for a multicomponent system), through
\begin{align}
  \mathbf{C}=\mathbf{K}^{-1}=\begin{pmatrix}
C_{1,1} & C_{1,2} & \cdots & C_{1,N}\\
C_{2,1} & \ddots & \ddots & \vdots\\
\vdots & \ddots & \ddots & \vdots\\
C_{N,1} & \cdots &\cdots & C_{N,N}\\
\end{pmatrix}.\label{chern}
\end{align}
Here the symmetric properties of matrix elements $C_{\sigma,\sigma'}=C_{\sigma',\sigma}$, and the off-diagonal part $C_{\sigma,\sigma'}$ is related to the drag Hall conductance between particles of the $\sigma$-th component and particles of  the $\sigma'$-th component. The total charge Hall conductance is given by~\cite{Wen1995}
\begin{align}
  \sigma_H=\mathbf{q}^{T}\cdot\mathbf{K}^{-1}\cdot\mathbf{q}
\end{align}
where $\mathbf{q}$ is the charge vector. For our SU$(N)$ symmetric systems, the $N$-component charge vector is $\mathbf{q}^T=[1,1,\cdots,1]$, and the total charge Hall conductance
\begin{align}
  \sigma_H=\sum_{i,j}C_{i,j}=\nu.
\end{align}
To uncover the topological nature of the FQH systems we numerically extract the Chern number matrix. Here, we utilize the scheme proposed in Refs.~\cite{Sheng2003,Sheng2006}.
With twisted boundary conditions $\psi(\cdots,\rr_{\sigma}^{i}+N_{\alpha}{\hat e_{\alpha}},\cdots)=\psi(\cdots,\rr_{\sigma}^{i},\cdots)\exp(i\theta_{\sigma}^{\alpha})$ where $\theta_{\sigma}^{\alpha}$ is the twisted angle for particles of the $\sigma$-th component in the $\alpha$-direction. In two-parameter $(\theta_{\sigma}^{x},\theta_{\sigma'}^{y})$ plane, the many-body Chern number $C_{\sigma,\sigma'}$ of the ground state wavefunction $\psi$ is defined as~\cite{Sheng2003,Sheng2006}
\begin{align}
  C_{\sigma,\sigma'}=\frac{1}{2\pi}\int d\theta_{\sigma}^{x}d\theta_{\sigma'}^{y}F_{\sigma,\sigma'}^{xy},\label{chern1}
\end{align}
with the Berry curvature
\begin{align}
    F_{\sigma,\sigma'}^{xy}=\mathbf{Im}\left(\langle{\frac{\partial\psi}{\partial\theta_{\sigma}^x}}|{\frac{\partial\psi}{\partial\theta_{\sigma'}^y}}\rangle
-\langle{\frac{\partial\psi}{\partial\theta_{\sigma'}^y}}|{\frac{\partial\psi}{\partial\theta_{\sigma}^x}}\rangle\right).
\end{align}
Similarly, one can define the spin-dependent twisted angles as $(\theta_{\sigma'}^{x},\theta_{\sigma'}^{y})=(-\theta_{\sigma}^{x},-\theta_{\sigma}^{y})$, representing the opposite propagation between particles of the $\sigma$-component and particles of the $\sigma'$-component, and obtain the quantized spin Chern number~\cite{Zeng2017}.

For $N=1$, the many-body ground state is the Laughlin $1/2$-FQH state in the lattice version at $\nu=1/2$ for hardcore bosons, while it is the Laughlin $1/3$-FQH state in the lattice version at $\nu=1/3$ for spinless fermion with large nearest-neighbor interaction $V/t\gg1$, as discussed extensively in many numerical works~\cite{BL2013}. The corresponding $\mathbf{K}$ matrix is just equal to $\mathbf{K}=N+1$ for bosons ($\mathbf{K}=2N+1$ for fermions), and there is only one chiral mode in the entanglement spectrum.

For $N=2$, the two-component variational wavefunction $\psi\propto\prod_{i<j}(z^{1}_i-z^{1}_j)^m(z^{2}_i-z^{2}_j)^m\prod_{i,j}(z^{1}_i-z^{2}_j)^n$ for FQH states with $\mathbf{K}=\begin{pmatrix}
m & n\\
n & m\\
\end{pmatrix}$ was firstly constructed by Halperin in Ref.~\cite{Halperin1983}, and now known as Halperin $(m,m,n)$ states. In Ref.~\cite{Zeng2017}, it has been numerically shown that for two-component hardcore bosons at filling factors $\nu=2/3$ with large SU$(2)$ invariant interactions $U/t\gg1,V=0$, the ground states are indeed the Halperin $(221)$ FQH states in the lattice version, classified by features including the unique Chern number matrix (inverse of $\mathbf{K}=\begin{pmatrix}
2 & 1\\
1 & 2\\
\end{pmatrix}$ matrix), the fractional charge and spin pumpings, and two parallel propagating edge modes. However for two-component fermions at $\nu=2/5$ FQH state, the numerical evidence of its $\mathbf{K}$ matrix classification is still desired, which will also be addressed in the Section~\ref{f332}.

For generic $N>2$, as a generalization from $N=1,2$ cases, the mathematical formula of repulsive interaction potentials between any two particles is unchanged. For Laughlin and Halperin wavefunctions, the power-law correlation $(z^{\sigma}_i-z^{\sigma'}_j)^m$ minimizes the two-body interaction energy (Haldane pseudopotential) between $\sigma$-component particle and $\sigma'$-component particle, as demonstrated in Landau levels~\cite{Laughlin1983,Haldane1983}. Thus in order to minimize the SU$(N)$ symmetric interaction energy in the flat bands, the correlated many-body wavefunction can be captured by $\psi\propto\prod_{\sigma,i<j}(z^{\sigma}_i-z^{\sigma}_j)^m\prod_{\sigma<\sigma',i,j}(z^{\sigma}_i-z^{\sigma'}_j)^n$, where the integer values $m,n$ are the diagonal and off-diagonal elements of $\mathbf{K}$ matrix in Chern-Simons theory~\cite{Wen1992a} (in equivalence to $\mathbf{G}$ matrix proposed in Ref.~\cite{Read1990}), and we conjecture that the $N\times N$-matrix for $N$-component bosonic FQH states with strong onsite Hubbard interactions $U/t\gg1,V=0$ at filling $\nu=N/(N+1)$
\begin{align}
  \mathbf{K}=\begin{pmatrix}
2 & 1 & \cdots & 1\\
1 & 2 & \ddots & \vdots\\
\vdots & \ddots & \ddots & \vdots\\
1 & \cdots &\cdots & 2\\
\end{pmatrix},\label{bkmatrix}
\end{align}
where the diagonal elements $\mathbf{K}_{j,j}=2$, and the off-diagonal elements $\mathbf{K}_{j,j'}=1$. Then,
\begin{align}
  \mathbf{C}=\frac{1}{N+1}\begin{pmatrix}
N & -1 & \cdots & -1\\
-1 & N & \ddots & \vdots\\
\vdots & \ddots & \ddots & \vdots\\
-1 & \cdots &\cdots & N\\
\end{pmatrix},\label{bcmatrix}
\end{align}
Similarly, we can write down the $N\times N$-matrix for $N$-component fermionic FQH states with strong nearest-neighbor interactions $V/t\gg1$ at filling $\nu=N/(2N+1)$
\begin{align}
  \mathbf{K}=\begin{pmatrix}
3 & 2 & \cdots & 2\\
2 & 3 & \ddots & \vdots\\
\vdots & \ddots & \ddots & \vdots\\
2 & \cdots &\cdots & 3\\
\end{pmatrix},\label{fkmatrix}
\end{align}
where the diagonal elements $\mathbf{K}_{j,j}=3$, and the off-diagonal elements $\mathbf{K}_{j,j'}=2$. From Eqs.~\ref{bkmatrix} and~\ref{fkmatrix}, we can easily obtain the determinants $\det\mathbf{K}=N+1$ for bosons and $\det\mathbf{K}=2N+1$ for fermions, which characterize the ground state degeneracy. From the inverse of $\mathbf{K}$ matrix, we derive its Chern number matrix, and verify that the total charge Hall conductance $\sigma_H=\nu$ for both bosons and fermions. Both of $\mathbf{K}$ matrices in Eqs.~\ref{bkmatrix} and~\ref{fkmatrix} are related to the Cartan matrix of the Lie algebra SU$(N)$ by a special linear group SL($N,\mathbb{Z}$) transformation, indicating that these ground FQH states exhibit an SU$(N)$ Kac-Moody symmetry at level one~\cite{Zee1991,Wen1992a}. Further the average particle filling of the $\sigma$-component in Eqs.~\ref{bkmatrix} and~\ref{fkmatrix} equals toto a constant $\nu_{\sigma}=\nu/N$, and these FQH states belong to the SU$(N)$ spin-singlet states.

For $N>2$, so far there are no numerical studies for $\mathbf{K}$ matrix of such FQH states due to its numerical difficulty in both ED and DMRG, where simulations beyond the ground state properties are needed. In the Section~\ref{b332}, we will discuss SU$(N)$ bosonic FQH states at $\nu=N/(N+1)$. From the Chern number matrix Eq.~\ref{chern}, one can also calculate the different Hall responses of the $\sigma$-th component in the $x$-direction under the insertion of flux quantum $\theta_{\sigma'}^{y}$ of the $\sigma'$-th component in the $y$-direction, which are related to the charge pumping.

\subsection{SU$(2)$ Halperin $(m,m,m-1)$ states}\label{f332}

\begin{figure}[t]
  \includegraphics[height=1.6in,width=3.4in]{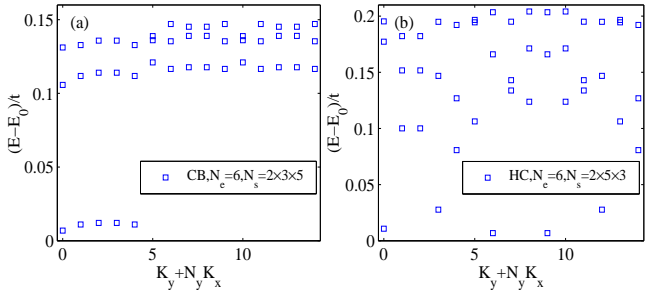}
  \caption{\label{energy2}(Color online) Numerical ED results for the low energy spectrum of two-component fermions $\nu=2/5,N_s=2\times3\times5,U=0,V=10t$ on two typical topological lattices: (a) $\pi$-flux checkerboard and (b) Haldane-honeycomb lattices, respectively.}
\end{figure}

In this section, we begin to discuss the numerical evidences of SU$(2)$ FQH states at given filling $\nu=2/5$ for fermions. First, in Figs.~\ref{energy2}(a) and~\ref{energy2}(b), we show the energy spectrum of several systems in strong interacting regime $U=0,V=10t$ at $\nu=2/5$ for two-component fermions. The key feature is that, there exists a five-fold quasidegenerate ground state manifold separated from higher-energy levels by a robust gap. For $U\gg t$, this degeneracy persists under the SU$(2)$ interaction. We also calculate the density and (pseudo)spin structure factors for the ground states, and exclude any possible charge or spin density wave orders as the competing ground states, due to the absence of the Bragg peaks in the results (we check them up to $N_s=2\times4\times5$ sites using DMRG with periodic boundary conditions).

\begin{figure}[t]
  \includegraphics[height=2.6in,width=3.4in]{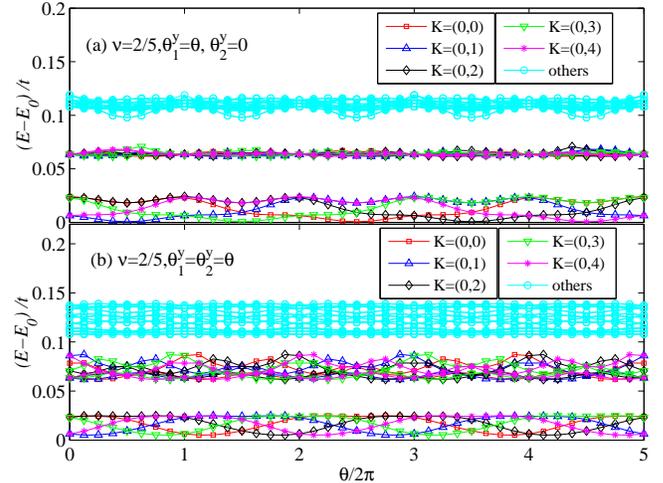}
  \caption{\label{flux}(Color online) Numerical ED results for the y-direction spectral flow of two-component fermionic systems $N_e=6,N_s=2\times3\times5$ at $U=10t,V=10t$ on the CB lattice under different insertion of flux quantum: (a) $\theta_{1}^{\alpha}=\theta,\theta_{2}^{\alpha}=0$; (b) $\theta_{1}^{\alpha}=\theta_{2}^{\alpha}=\theta$.}
\end{figure}

Next, we plot the low energy spectra under the variation of $\theta_{\sigma}^{\alpha}$. As shown in Figs.~\ref{flux}(a) and~\ref{flux}(b), these Abelian ground states evolve into each other without mixing with the higher levels. Interestingly, for two-component fermions at $\nu=2/5$, the system evolves back to itself after the insertion of five flux quanta for both $\theta_{1}^{\alpha}=\theta_{2}^{\alpha}=\theta$, and $\theta_{1}^{\alpha}=\theta,\theta_{2}^{\alpha}=0$, indicating its $1/5$ fractional quantization of quasiparticles.

\begin{figure}[t]
  \includegraphics[height=1.5in,width=3.4in]{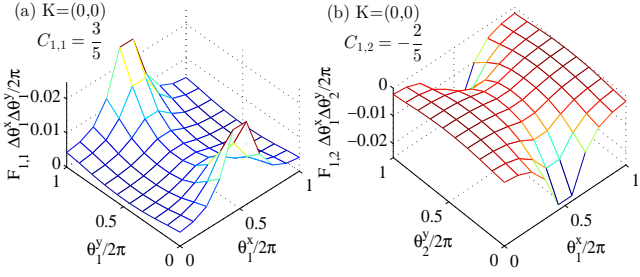}
  \caption{\label{berry}(Color online) Berry curvatures $F_{\sigma,\sigma'}^{xy}\Delta\theta_{\sigma}^{x}\Delta\theta_{\sigma'}^{y}/2\pi$ for the $K=(0,0)$ ground state of two-component fermionic systems $N_e=6,N_s=2\times3\times5$ at $U=10t,V=10t$ on the CB lattice in the parameter plane: (a) $(\theta_{1}^{x},\theta_{1}^{y})$ and (b) $(\theta_{1}^{x},\theta_{2}^{y})$.}
\end{figure}

For the five ground states with $K=(0,i), (i=0,1,2,3,4)$ of two-component fermions at $N_e=6,N_s=2\times3\times5$, by numerically calculating the Berry curvatures using $m\times m$ mesh squares in the boundary phase space with $m\geq10$ we obtain $\sum_{i=1}^{5} C_{1,1}^i=3$, and $\sum_{i=1}^{5} C_{1,2}^i=-2$. For single ground state at $K=(0,0)$, we plot the corresponding Berry curvature $F_{1,1},F_{1,2}$ in Figs.~\ref{berry}(a) and~\ref{berry}(b), which vary smoothly. All of the above imply a $2\times2$ $\mathbf{C}$ matrix, namely,
\begin{align}
  \mathbf{C}=\begin{pmatrix}
C_{1,1} & C_{1,2}\\
C_{2,1} & C_{2,2}\\
\end{pmatrix}=\frac{1}{5}\begin{pmatrix}
3 & -2\\
-2 & 3\\
\end{pmatrix}.
\end{align}
Thus we can obtain the $\mathbf{K}$ matrix from the inverse of the $\mathbf{C}$ matrix, namely $\mathbf{K}=\mathbf{C}^{-1}=\begin{pmatrix}
3 & 2\\
2 & 3\\
\end{pmatrix}$. Therefore, we establish that five-fold ground states for two-component fermion at $\nu=2/5$ are indeed Halperin (332) states in lattice version, and the five-fold degeneracy coincides with the determinant $\det\mathbf{K}$, as predicted in Ref.~\cite{Wen1995}.

\subsection{SU$(N)$ FQH states with $N>2$}\label{b332}

\begin{figure}[b]
  \includegraphics[height=1.34in,width=3.4in]{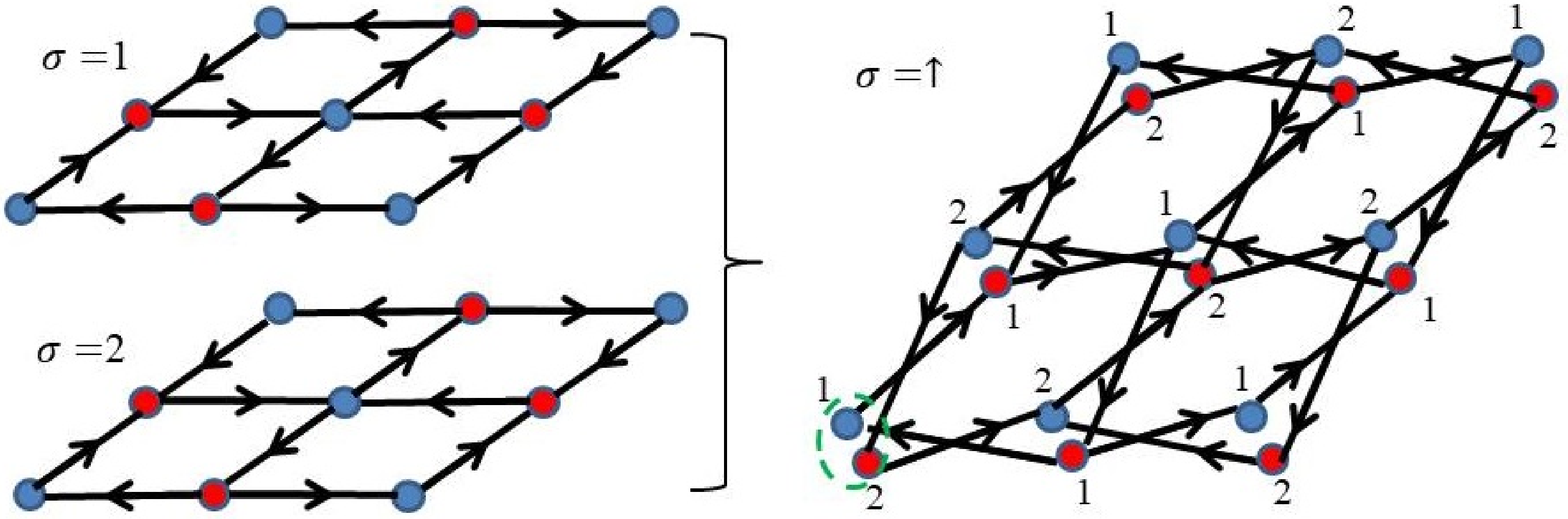}
  \caption{\label{lattice2}(Color online) The square lattice model with Chern number two engineered from two decoupled $\pi$-flux checkerboard lattices $\sigma=1,2$. The arrow directions present the signs of the phases $\phi$ in the nearest-neighboring hopping terms. The small dashed green cycle indicate the new unit cell in the square lattice, denoted as spin-$\uparrow$.}
\end{figure}

Following the last section, we move on to discuss the emergence of SU$(N)$ bosonic FQH states at $\nu=N/(N+1)$ with strong onsite Hubbard repulsion $U$. Here we focus on the $N=4$ case, and leave the discussion about $N=3$ case into the appendix. For the $N=4$ multicomponent systems, in order to overcome the numerical difficulty, we stack the two-component topological CB lattices $H_{CB}^{\sigma},\sigma=1,2$ in Eq.~\ref{cbl} into one equivalent single-layer square lattice with Chern number two, as engineered in Ref.~\cite{Yang2012}. Similarly, we stack the other two-component CB lattices $H_{CB}^{\sigma},\sigma=3,4$ in Eq.~\ref{cbl} into another equivalent single-layer square lattice with Chern number two. Under this construction, the hopping parameters are kept unchanged, but the Chern number of the lowest band changes from one to two. Now we relabel the particle on the first (second) square lattice layer as
spin-$\uparrow$ (spin-$\downarrow$), as indicated in Fig.~\ref{lattice2}. By introducing the new twisted boundary conditions on the square lattice $\psi(\cdots,\rr_{\sigma}^{i}+N_{\alpha}{\hat e_{\alpha}},\cdots)=\psi(\cdots,\rr_{\sigma}^{i},\cdots)\exp(i\theta_{\sigma}^{\alpha})$ where $\theta_{\sigma}^{\alpha}$ is the twisted angle for particles of spin-$\sigma$ in the $\alpha$-direction, we define a new many-body Chern number $C_{\sigma,\sigma'}$ in analogous to Eq.~\ref{chern1}. Therefore the $4\times4$ Chern number matrix at filling $\nu=4/5$ of the lowest flat band with Chern number one in Eq.~\ref{chern} is reduced to a new $2\times2$ matrix at filling $\widetilde{\nu}=(N_{\uparrow}+N_{\downarrow})/N_s=2/5$ of the lowest flat band with Chern number two,
\begin{align}
  \widetilde{\mathbf{C}}=\begin{pmatrix}
C_{\uparrow,\uparrow} & C_{\uparrow,\downarrow}\\
C_{\downarrow,\uparrow} & C_{\downarrow,\downarrow}\\
\end{pmatrix}
\end{align}
with the relationship
\begin{align}
&C_{\uparrow,\uparrow}=C_{1,1}+C_{1,2}+C_{2,1}+C_{2,2},\label{su41}\\
&C_{\uparrow,\downarrow}=C_{1,3}+C_{1,4}+C_{2,3}+C_{2,4},\label{su42}\\
&C_{\downarrow,\uparrow}=C_{3,1}+C_{3,2}+C_{4,1}+C_{4,2},\\
&C_{\downarrow,\downarrow}=C_{3,3}+C_{3,4}+C_{4,3}+C_{4,4}.
\end{align}

\begin{figure}[t]
  \includegraphics[height=1.5in,width=3.4in]{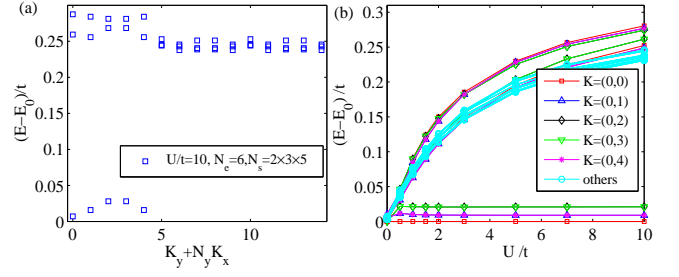}
  \caption{\label{energy4}(Color online) Numerical ED results for the low energy spectrum of two-component bosons $\widetilde{\nu}=2/5,N_s=2\times3\times5,V=0$ on the square lattice with Chern number two: (a) at $U=10t$ and (b) with increasing $U$.}
\end{figure}

\begin{figure}[t]
  \includegraphics[height=2.8in,width=3.4in]{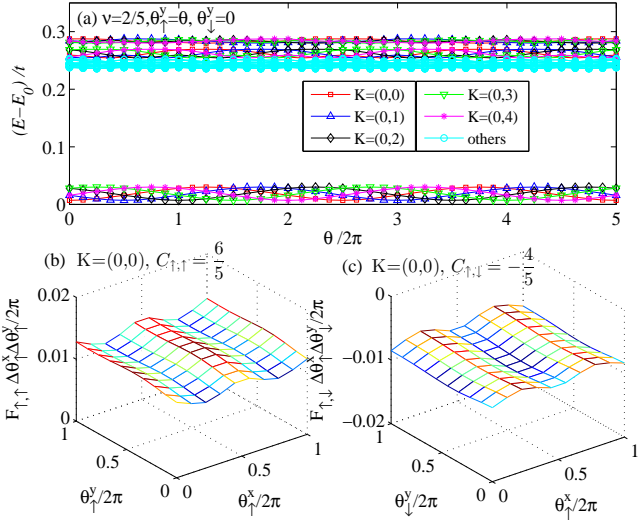}
  \caption{\label{berry4}(Color online) (a) Numerical ED results for the y-direction spectral flow of two-component bosonic systems $\widetilde{\nu}=2/5,N_s=2\times3\times5$ on the square lattice with Chern number two at $U=10t,V=0$. Berry curvatures $F_{\sigma,\sigma'}^{xy}\Delta\theta_{\sigma}^{x}\Delta\theta_{\sigma'}^{y}/2\pi$ for the $K=(0,0)$ ground state of two-component bosonic systems $N_e=6,N_s=2\times3\times5$ on the square lattice with Chern number two at $U=10t,V=0$ in the parameter plane: (a) $(\theta_{\uparrow}^{x},\theta_{\uparrow}^{y})$ and (b) $(\theta_{\uparrow}^{x},\theta_{\downarrow}^{y})$.}
\end{figure}

In the ED study, we show the energy spectrum of strongly interacting effective two-component bosons at $\widetilde{\nu}=2/5$ in Figs.~\ref{energy4}(a) and~\ref{energy4}(b), where the five-fold quasidegenerate ground state manifold is protected by a robust gap. As shown in Fig.~\ref{berry4}(a), these ground states evolve into each other without mixing with the higher levels, and the energy recovers itself after the insertion of five flux quanta for $\theta_{\uparrow}^{\alpha}=\theta,\theta_{\downarrow}^{\alpha}=0$, indicating its $1/5$ fractional quantization of quasiparticles.

For the five ground states with $K=(0,i),(i=0,1,2,3,4)$, by numerically calculating the Berry curvatures using $m\times m$ mesh squares in the boundary phase space with $m\geq10$ we obtain $\sum_{i=1}^{5} C_{\uparrow,\uparrow}^i=6$, and $\sum_{i=1}^{5} C_{\uparrow,\downarrow}^i=-4$. For single ground state at $K=(0,0)$, we plot the corresponding Berry curvature $F_{\uparrow,\uparrow},F_{\uparrow,\downarrow}$ in Figs.~\ref{berry}(a) and~\ref{berry}(b), which vary very smoothly over the plane. All of the above imply a $2\times2$ $\widetilde{\mathbf{C}}$ matrix, namely,
\begin{align}
  \widetilde{\mathbf{C}}=\begin{pmatrix}
C_{\uparrow,\uparrow} & C_{\uparrow,\downarrow}\\
C_{\downarrow,\uparrow} & C_{\downarrow,\downarrow}\\
\end{pmatrix}=\frac{1}{5}\begin{pmatrix}
6 & -4\\
-4 & 6\\
\end{pmatrix}.\label{su4}
\end{align}
Due to the permutation symmetry of the original Hamiltonian Eq.~\ref{cbl}, one should have the properties of the ground state $C_{j,j}=C_{1,1},C_{j,j'}=C_{1,2}$. Thus from both Eq.~\ref{su41} and Eq.~\ref{su42}, we can derive the values $C_{1,1}=4/5,C_{1,2}=-1/5$, demonstrating the Chern number matrix governed by Eq.~\ref{bcmatrix}. Therefore, we establish that five-fold ground states for four-component hardcore bosons at $\nu=4/5$ are indeed SU$(N=4)$ FQH states classified by the $\mathbf{K}$ matrix Eq.~\ref{bkmatrix} in the lattice version, and the five-fold degeneracy coincides with the determinant $\det\mathbf{K}$.

\section{Fractional charge pumpings}\label{pumping}

In this section, we further discuss the corresponding charge (spin) pumpings complementary to the Chern number matrix for SU$(N)$ FQH states obtained in Sec.~\ref{kmatrix}. In DMRG, we calculate the charge pumping of the ground state under the insertion of flux quantum on large cylinder systems~\cite{Gong2014}, in connection to the quantized Hall conductance. In DMRG we partition  the cylinder along the $x$-direction into two halves with equal lattice sites. By inserting one flux quantum $\theta_{\sigma}^{y}=\theta,\theta_{\sigma'}^{y}=0$ from $\theta=0$ to $\theta=2\pi$ on the cylinder system, the expectation value of the particle number of the $\sigma$-th component  on the left side equals to $N_{\sigma}^{L}(\theta)=tr[\widehat{\rho}_L(\theta)\widehat{N}_{\sigma}^{L}]$, where $N_{\sigma}^{L}$ is the particle number of spin-$\sigma$ in the left cylinder part, and $\widehat{\rho}_L$ the reduced density matrix of the corresponding left part.  The net charge transfer of the $\sigma$-th component particle from the right side to the left side during each cycle is encoded by
\begin{align}
  \Delta N_{\sigma}=N_{\sigma}^{L}(2\pi)-N_{\sigma}^{L}(0)=C_{\sigma,\sigma}.
\end{align}

\begin{figure}[t]
  \includegraphics[height=1.6in,width=3.4in]{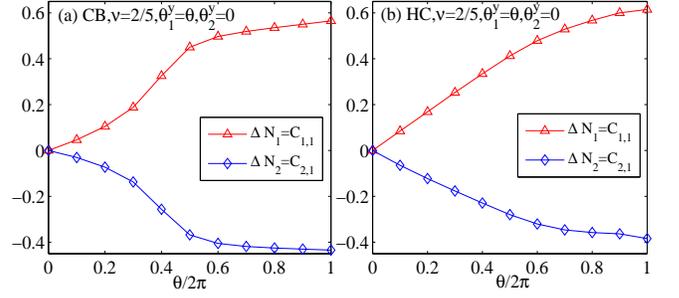}
  \caption{\label{pump}(Color online) The charge transfer for two-component fermions on the $N_y=3$ cylinder at $\nu=2/5,U=10t,V=10t$ with inserting flux $\theta_{1}^y=\theta,\theta_{2}^y=0$: (a) $\pi$-flux checkerboard lattice; (b) Haldane-honeycomb lattice. Here the calculation is performed using finite DMRG with cylinder length $L_x=N_x=30$ and the maximal kept number of states 2400.}
\end{figure}

\begin{figure}[b]
  \includegraphics[height=1.6in,width=2.6in]{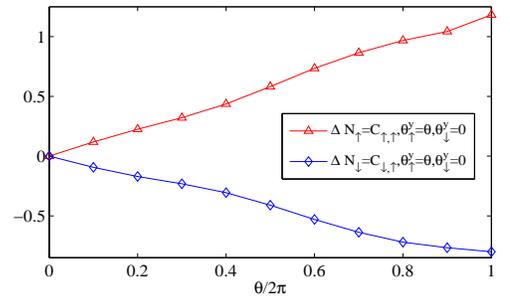}
  \caption{\label{pump4}(Color online) The charge transfer for two-component bosons on the $N_y=3$ cylinder of the square lattice with Chern number two at $\widetilde{\nu}=2/5,U=10t,V=0$ with inserting flux $\theta_{\uparrow}^y=\theta,\theta_{\downarrow}^y=0$. Here the calculation is performed using finite DMRG with cylinder length $L_x=N_x=30$ and the maximal kept number of states 1600.}
\end{figure}

Similarly, due to the drag Hall conductance induced by $C_{\sigma',\sigma}$, there also should be the charge transfer among different components, namely
\begin{align}
  \Delta N_{\sigma'}=N_{\sigma'}^{L}(2\pi)-N_{\sigma'}^{L}(0)=C_{\sigma',\sigma}.
\end{align}
As shown in Figs.~\ref{pump}(a) and~\ref{pump}(b), for two-component fermions at $\nu=2/5$, a fractional charge $\Delta N_1\simeq0.6$ is pumped in one species, and a fractional charge $\Delta N_2\simeq-0.4$ pumped in the other species by threading one flux quantum $\theta_{1}^y=\theta,\theta_{2}^y=0$. By threading one flux quantum $\theta_{1}=\theta_{2}=\theta$ in both species, the total charge pump is just given by $\Delta N=\sum_{\sigma,\sigma'}C_{\sigma,\sigma'}=\nu=\sigma_H$.

Similarly, for effective two-component bosons at $\widetilde{\nu}=2/5$ on the square lattice with Chern number two, by threading one flux quantum $\theta_{\uparrow}^y=\theta,\theta_{\downarrow}^y=0$, a fractional charge $\Delta N_{\uparrow}\simeq1.2$ is pumped in one species, and a fractional charge $\Delta N_{\downarrow}\simeq-0.8$ pumped in the other species, as indicated in Fig.~\ref{pump4}.

\section{Relationship to topological bands with Chern number $N$}\label{fci}

In this section, we study the possible relationship of our constructed multicomponent FQH states at $\nu=N/(MN+1),M=1,2$ in Sec.~\ref{kmatrix} to the single-component FQH states in topological flat bands with Chern number $N$. For hardcore bosons at one-third filling of the flat band with Chern number $C=2$, it hosts three-fold degenerate ground states reminiscent of two-component Halperin (221) states.

For $N>2$, following the methods of Ref.~\cite{Yang2012}, we twist the $N$-layer checkerboard lattices in Eq.~\ref{cbl} into single layer square lattice, where the lowest flat band has the Chern number $C=N$. Now each unit cell contains $N$ inequivalent sites from different $N$ layers. When SU$(N)$ interactions in Eq.~\ref{interact} between different components are considered, we expect the emergence of a series of FQH states at fillings $\widetilde{\nu}=1/(N+1)$ for bosons ($\widetilde{\nu}=1/(2N+1)$ for fermions) on the square lattice with SU$(N)$-invariant interaction,
\begin{align}
&H^{[C=N]}=\sum_{\rr}\sum_{l=1}^{N}\left\{ t_{1}\left(C_{\rr+{\hat e_{x}}}^{[l+1]\dagger}+e^{i2l\phi}C_{\rr-{\hat e_{y}}}^{[l+1]\dagger}\right)C_{\rr}^{[l]}\right.\notag \\
 &\quad+t_{2}\left[e^{-i\left(2l-1\right)\phi}C_{\rr+{\hat e_{x}}+{\hat e_{y}}}^{[l]\dagger}+e^{i\left(2l-1\right)\phi}C_{\rr-{\hat e_{x}}-{\hat e_{y}}}^{[l]\dagger}\right. \notag \\
&\quad\left.\left.+e^{i\left(2l+1\right)\phi}C_{\rr+{\hat e_{x}}-{\hat e_{y}}}^{[l+2]\dagger}\right]C_{\rr}^{[l]}+\mathrm{H.c.}\right\}+ V_{int},\\
&\quad\quad V_{int}=U\sum_{l\neq l'}\sum_{\rr}n_{\rr}^{l}n_{\rr}^{l'}+V\sum_{l,l'}\sum_{\langle\rr,\rr'\rangle}n_{\rr}^{l}n_{\rr}^{l'},
\end{align}
where $l=1,\cdots,N$ denote the inequivalent sites in each unit cell from different layers, and $n_{\rr}^{l}=C_{\rr}^{[l]\dagger}C_{\rr}^{[l]}$ the particle operator.

\begin{figure}[t]
  \includegraphics[height=1.55in,width=3.4in]{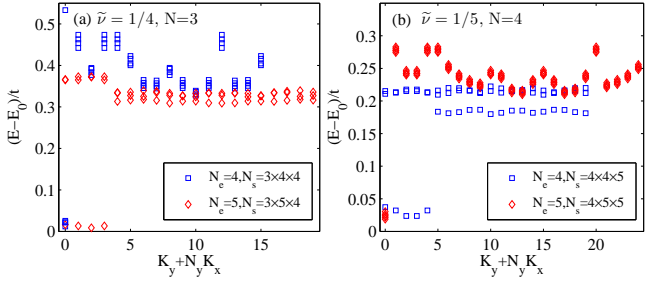}
  \caption{\label{energyc3}(Color online) Numerical ED results for the low energy spectrum of strongly interacting hardcore bosons at filling $\widetilde{\nu}=1/(N+1)$ on the topological square lattice with Chern number $N$: (a) $N=3$ and (b) $N=4$, respectively. The lattice parameters are the same as those in Ref.~\cite{Yang2012}, with $U=10t,V=0$.}
\end{figure}

In Figs.~\ref{energyc3}(a) and~\ref{energyc3}(b), we plot the low energy spectrum of strongly interacting hardcore bosons at filling $\widetilde{\nu}=1/(N+1)$ on the topological square lattice with Chern number $C=N$. Clearly, the ground states have $(N+1)$-fold degeneracy, and using the twisted boundaries we numerically verify that the many-body Chern number equals to the Hall conductance $N\widetilde{\nu}=\nu=\sigma_H$. Both the degeneracy and the Hall conductance match well with the predictions of the $\mathbf{K}$ matrix in Eq.~\ref{bkmatrix}. In Appendix~\ref{ces}, we show that the three branches of edge modes also match well with the three positive eigenvalues of the $\mathbf{K}$ matrix in Eq.~\ref{bkmatrix}.

\begin{figure}[t]
  \includegraphics[height=1.60in,width=3.4in]{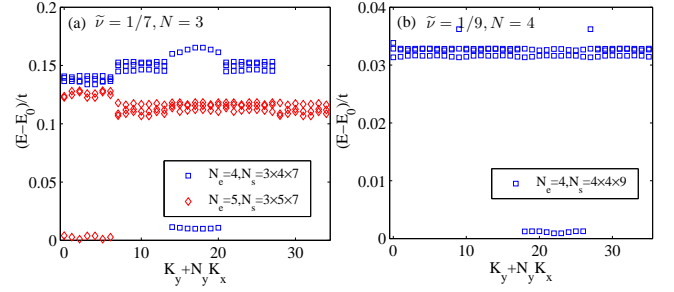}
  \caption{\label{energyf3}(Color online) Numerical ED results for the low energy spectrum of strongly interacting spinless fermions at filling $\widetilde{\nu}=1/(2N+1)$ on the topological square lattice with Chern number $N$: (a) $N=3$ and (b) $N=4$, respectively. The lattice parameters are the same as those in Ref.~\cite{Yang2012}, with $U=V=10t$.}
\end{figure}

Similarly, in Figs.~\ref{energyf3}(a) and~\ref{energyf3}(b), we plot the low energy spectrum of strongly interacting fermions at filling $\widetilde{\nu}=1/(2N+1)$ on the topological square lattice with Chern number $N=3$. Clearly, the ground states have $(2N+1)$-fold degeneracy, and using the twisted boundaries we numerically verify that the many-body Chern number equals to the Hall conductance $N\widetilde{\nu}=\nu=\sigma_H$. Both the degeneracy and the Hall conductance match well with the predictions of the $\mathbf{K}$ matrix in Eq.~\ref{fkmatrix}.

Combined with the results of FQH states for $N=1,2$, it is physically convincing to draw the one-to-one correspondence between the $N$-component FQH states at $\nu=N/(N+1)$ for hardcore bosons ($\nu=N/(2N+1)$ for fermions) on the topological lattice with Chern number one, and the single-component FQH states at $\widetilde{\nu}=1/(N+1)$ for hardcore bosons ($\widetilde{\nu}=1/(2N+1)$ for spinless fermion) on the topological lattice with Chern number $N$.

\section{Summary and Discussions}\label{summary}

In summary, we show that $N$-component hardcore bosons and fermions in topological lattice models could host SU$(N)$ FQH states at a partial filling $\nu=N/(MN+1)$ of the lowest Chern band ($M=1$ for bosons and $M=2$ for fermions), with fractional topological properties, including the ground state degeneracy and fractional charge pumpings, characterized by a $N\times N$ $\mathbf{K}$ matrix. The close relationship to the single component FQH states at fractional fillings $\nu=1/(MN+1)$ of the lowest Chern band ($M=1$ for bosons and $M=2$ for fermions) on the topological lattice models with high Chern number $N>2$ is also examined. We note that for $N=2$, our FQH states are indeed the spin-singlet states proposed in Ref.~\cite{Ardonne1999}, where the lower sequential FQH states with $\mathbf{K}=\begin{pmatrix}
4 & 3\\
3 & 4\\
\end{pmatrix}$ at filling $\nu=2/7$ for bosons is left for near future study.

\begin{acknowledgements}
We thank Duncan Haldane for stimulating discussions.
This work is supported by the U.S. Department of Energy, Office of Basic Energy Sciences under Grant No.
DE-FG02-06ER46305 (T.S.Z, and D.N.S.). 
D.N.S also acknowledges travel support from Princeton MRSEC through the National Science Foundation  Grant No. DMR-1420541.
\end{acknowledgements}

\appendix

\section{SU$(N=3)$ FQH states}\label{ces}

\begin{figure}[t]
  \includegraphics[height=1.6in,width=3.4in]{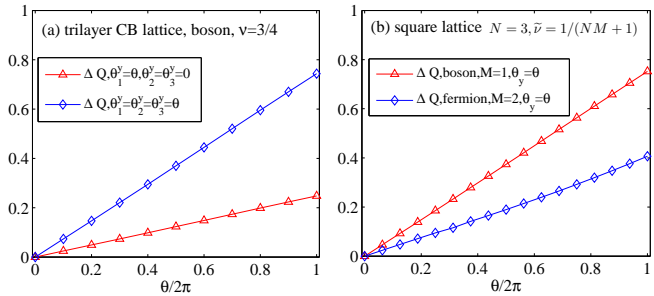}
  \caption{\label{pump3}(Color online) (a) The charge transfer for $N=3$-component hardcore bosons on the $N_y=4$ cylinder  of the $\pi$-flux checkerboard lattice with Chern number one at $\nu=3/4,U=10t,V=0$ under two different flux insertion methods $\theta_{1}^y=\theta,\theta_{2}^y=\theta_{3}^y=0$ and $\theta_{1}^y=\theta_{2}^y=\theta_{3}^y=\theta$; (b) The charge transfer for single-component particles on the $N_y=4$ cylinder of the square lattice with Chern number $N=3$ at $\widetilde{\nu}=1/(NM+1)$ under the flux insertion method $\theta^y=\theta$. Here the calculation is performed using infinite DMRG and the maximal kept number of states 1200.}
\end{figure}

In this appendix, we will discuss the topological signatures of SU$(N=3)$ FQH states for bosons and fermions. For bosonic trilayer $\pi$-flux checkerboard lattice models with small system size $N_e=6,N_s=2\times2\times4,\nu=3/4$ in Eq.~\ref{cbl}, our ED calculation gives indeed the four-fold degenerate ground states for strong interactions $U\gg t,V=0$. For large system sizes, we obtain the topological information from the fractional charge pumping and entanglement spectrum based on the infinite DMRG method. Analogous to Sec.~\ref{pumping}, by inserting one flux quantum from $\theta=0$ to $\theta=2\pi$ for cylinder systems, we define the total charge pumping from the right side to the left side as $\Delta Q=N^{L}(2\pi)-N^{L}(0)$ with $N^{L}(\theta)=\sum_{\sigma}tr[\widehat{\rho}_L(\theta)\widehat{N}_{\sigma}^{L}]$ the total charge in the left side.

Figure~\ref{pump3}(a) shows the total charge pumping $\Delta Q$ of strongly interacting hardcore bosons in trilayer $\pi$-flux checkerboard lattice under two different flux insertion methods (i) $\theta_{1}^y=\theta,\theta_{2}^y=\theta_{3}^y=0$ and (ii) $\theta_{1}^y=\theta_{2}^y=\theta_{3}^y=\theta$. When one flux quantum is inserted into only one of the layers $\theta_{1}^y=\theta,\theta_{2}^y=\theta_{3}^y=0$, the total charge pumping $\Delta Q=C_{1,1}+C_{2,1}+C_{3,1}=1/4$ from Eq.~\ref{bcmatrix}; When one flux quantum is inserted in all of the layers $\theta_{1}^y=\theta_{2}^y=\theta_{3}^y=\theta$, the total charge pumping $\Delta Q=\sum_{j,j'}C_{j,j'}=\nu=\sigma_H$ from Eq.~\ref{bcmatrix}. All the topological signatures are consistent with the $\mathbf{K}$ matrix in Eq.~\ref{bkmatrix}.

For strongly interacting single-component system at fractional fillings $\widetilde{\nu}=1/(NM+1)$ ($M=1$ for bosons and $M=2$ for fermions) on the topological square lattice with Chern number $C=N=3$, Figure~\ref{pump3}(b) shows the total charge pumping $\Delta Q$ under the flux insertion $\theta^y=\theta$. We obtain $\Delta Q=N\widetilde{\nu}=\nu=\sigma_H$, which is just the total charge pumping $\Delta Q=\sum_{j,j'}C_{j,j'}$ of $N$-component systems  under the flux insertion $\theta_{j}^y=\theta,j=1,\ldots,N$. For the square lattice with Chern number $N=3$, there are three degenerate edge modes differed by a momentum phase $2\pi/3$ in the Brillouin zone. By calculating the low-lying bulk entanglement spectrum (ES) of the ground state~\cite{Li2008}, we confirm that there exist three forward-moving chiral branches in the same direction in different charge sectors for bosonic fractional quantum Hall effect at filling $\widetilde{\nu}=1/4$, which are consistent with three positive eigenvalues of $\mathbf{K}$ matrix in Eq.~\ref{bkmatrix}.

\end{document}